\documentclass[11pt]{article}

\usepackage[preprint]{acl}

\usepackage{times}
\usepackage{latexsym}

\usepackage[T1]{fontenc}
\usepackage[utf8]{inputenc}

\usepackage{microtype}

\usepackage{inconsolata}

\usepackage{graphicx}
\usepackage{hyperref}

\title{Beyond Single Bugs: Benchmarking Large Language Models for Multi-Vulnerability Detection}

\author{
Chinmay Pushkar \\
BITS Pilani \\
\texttt{\small{f20210887@pilani.bits-pilani.ac.in}}
\And
Sanchit Kabra \\
Virginia Tech \\
\texttt{\small{sanchit23@vt.edu}}
\AND
Dhruv Kumar \\
BITS Pilani \\
\texttt{\small{dhruv.kumar@pilani.bits-pilani.ac.in}}
\And
Jagat Sesh Challa \\
BITS Pilani \\
\texttt{\small{jagatsesh@pilani.bits-pilani.ac.in}}
}

\begin{document}
\maketitle
\begin{abstract}
Large Language Models (LLMs) have demonstrated significant potential in automated software security, particularly in vulnerability detection. However, existing benchmarks primarily focus on isolated, single-vulnerability samples or function-level classification, failing to reflect the complexity of real-world software where multiple interacting vulnerabilities often coexist within large files. Recent studies indicate that LLMs suffer from "count bias" and "selection bias" in multi-label tasks, yet this has not been rigorously quantified in the domain of code security. In this work, we introduce a comprehensive benchmark for Multi-Vulnerability Detection across four major languages: C, C++, Python, and JavaScript. We construct a dataset of 40,000 files by systematically injecting controlled counts of vulnerabilities (1, 3, 5, and 9) into long-context code samples (7.5k–10k tokens) sourced from CodeParrot. We evaluate five state-of-the-art LLMs, including GPT-4o-mini, Llama-3.3-70B, and the Qwen-2.5 series.  Our results reveal a sharp degradation in performance as vulnerability density increases. While Llama-3.3-70B achieves near-perfect F1 scores (~0.97) on single-vulnerability C tasks, performance drops by up to 40\% in high-density settings. Notably, Python and JavaScript show distinct failure modes compared to C/C++, with models exhibiting severe "under-counting" (Recall dropping to <0.30) in complex Python files.
\end{abstract}

\section{Introduction}

The security of software infrastructure is paramount, yet manual code review remains unscalable. Large Language Models (LLMs) have emerged as powerful tools for static analysis, capable of understanding syntax and semantics to identify security flaws \cite{chen2021evaluatinglargelanguagemodels}. Unlike traditional rule-based solvers (e.g., CodeQL), LLMs can detect semantic vulnerabilities and subtle logic errors. Consequently, evaluating LLMs for vulnerability detection has become a central focus of software engineering research.

Despite their promise, current evaluations of LLMs in security are limited by the simplistic nature of existing benchmarks \cite{ahmed2025secvulevalbenchmarkingllmsrealworld}. Datasets such as \href{https://huggingface.co/datasets/DetectVul/devign}{Devign} \cite{zhou2019devigneffectivevulnerabilityidentification} or \href{https://huggingface.co/datasets/google/reveal}{Reveal} \cite{jacovi2024chainofthoughtstrongweakestlink} typically treat vulnerability detection as a binary classification task (Vulnerable vs. Safe) or focus on identifying a single bug within a small function. This contradicts real-world scenarios where source files are large, complex, and often contain multiple, distinct vulnerabilities (e.g., a file containing both a buffer overflow and a command injection).

Prior work has established benchmarks like \href{https://huggingface.co/datasets/openai/openai_humaneval}{HumanEval} \cite{chen2021evaluatinglargelanguagemodels} or \href{https://huggingface.co/datasets/Muennighoff/mbpp}{MBPP} \cite{austin2021programsynthesislargelanguage} for functional correctness, and CyberSecEval \cite{bhatt2024cyberseceval} for safety alignment. However, these do not assess the model's ability to exhaustively enumerate vulnerabilities. A recent study \cite{xu2025satabenchselectapplybenchmark} highlighted that LLMs struggle with "selection bias" (favoring specific answers regardless of content) and "count bias" (failing to predict the correct number of answers) in multiple-choice tasks. A significant research gap exists in applying this insight to code security: We do not know if LLMs can reliably detect all vulnerabilities in a file, or if they stop after finding the most obvious one.

To address this gap, we propose a systematic stress-test for Multi-Vulnerability Detection. We move beyond binary classification to a multi-label extraction task. We introduce a pipeline that utilizes an Oracle LLM to identify feasible injection points in clean code and systematically injects specific combinations of the \href{https://cwe.mitre.org/top25/archive/2024/2024_top25_list}{2024 Top-25 CWEs} (Common Weakness Enumerations). This allows us to control the ground truth density (1, 3, 5, or 9 vulnerabilities per file) and evaluate model robustness against count bias in the context of code vulnerability.

We construct a dataset from the CodeParrot GitHub Code \cite{dataset} corpus covering C, C++, Python, and JavaScript, focusing on file-level code samples of 7,500–10,000 tokens. Prior work shows that LLMs experience substantial retrieval degradation well before their maximum context limits due to positional bias and the “Lost in the Middle” effect \cite{liu2023lostmiddlelanguagemodels}. This context range therefore provides a realistic and challenging setting that stresses long-range attention without introducing truncation artifacts. Using a hybrid injection strategy, we generate five dataset variants per language—Clean, 1-Vuln, 3-Vuln, 5-Vuln, and 9-Vuln.

We evaluated five models: Qwen2.5-32B-Instruct, Qwen2.5-72B-Instruct \cite{qwen2.5}, Llama-3.3-70B-Instruct \cite{touvron2023llamaopenefficientfoundation}, Mistral-Small-3.2-24B \cite{mistral_small}, and GPT-4o-mini \cite{gpt4o_mini}. We utilized metrics including Precision, Recall, F1-Score, and a novel "ExactFile" metric (percentage of files where the model identified the exact set of vulnerabilities perfectly) to measure performance.

Our experiments show that while modern LLMs perform strongly in single-vulnerability detection, their reliability degrades sharply as vulnerability density increases. In C++, Llama-3.3-70B achieves an F1 score of 0.90 on single-vulnerability files, but this drops to 0.62 for files containing nine vulnerabilities, accompanied by a substantial decline in Recall. Across all languages, higher vulnerability counts lead to systematic under-reporting, reflected in steadily increasing Mean Absolute Error (MAE) values (+209.3\% for Llama-3.3-70B for C++) and collapsing Recall values (-45.3\% for Llama-3.3-70B for C++). Although Llama-3.3-70B consistently outperforms other models, even it suffers a 30–40\% reduction in ExactFile accuracy when moving from single- to high-density vulnerability settings, highlighting a fundamental limitation in exhaustive, file-level vulnerability detection.

\section{Related Work}

Research on automated vulnerability detection spans curated datasets, classical and learning-based analysis techniques, and, more recently, the use of large language models (LLMs). Our work lies at the intersection of these areas, addressing gaps in multi-vulnerability, file-level evaluation of generative models.

\textbf{Vulnerability Detection Datasets.}
Early work has depended on structured datasets such as the NIST Juliet Test Suite \cite{6329885}, which offers extensive CWE coverage across C/C++ and Java but consists of synthetic, simplified examples that lack real-world complexity. To address this limitation, datasets like Devign \cite{zhou2019devigneffectivevulnerabilityidentification}, Big-Vul \cite{10.1145/3379597.3387501}, and Reveal \cite{jacovi2024chainofthoughtstrongweakestlink} mined vulnerabilities from open-source repositories, producing more realistic samples. However, these datasets are typically function-level and support only binary vulnerability classification, limiting their ability to model multiple issues within the same file. As a result, they do not adequately test a model’s ability to identify, classify, and count co-located vulnerabilities. Our work diverges by synthetically generating controlled, file-level samples with multiple vulnerabilities, enabling evaluation of detection performance under complex distributions.

\textbf{LLMs for Code Security.}
Recent years have seen growing interest in applying LLMs—such as GPT-4, Llama, and Mistral—to tasks including static analysis, vulnerability explanation, and patch suggestion. Despite promising results, most evaluations still rely on traditional single-vulnerability datasets or binary classification tasks. This restricts insight into whether models can perform a holistic security audit, particularly when a file may contain multiple issues of different CWE types. Our work addresses this gap by framing vulnerability detection as a generative extraction task at the file level.

\textbf{Evaluation Challenges in LLMs.}
Evaluating LLMs on multi-output tasks remains difficult. Studies such as \cite{itzhak2024instructedbiasinstructiontunedlanguage} highlight instruction-tuning biases that cause models to output incorrect item counts or favor frequent labels. Related work on selection bias \cite{xu2025satabenchselectapplybenchmark} and long-context weaknesses—most notably the “Lost in the Middle” phenomenon \cite{liu2023lostmiddlelanguagemodels} further suggests systematic retrieval failures. These limitations are especially problematic for vulnerability detection, where under-reporting or misclassification directly affects practical usability. Our benchmark explicitly quantifies these effects through metrics such as MAE and ExactFile.

\textbf{LLM Benchmarking in Code.}
Popular benchmarks like HumanEval \cite{chen2021evaluatinglargelanguagemodels}, MBPP \cite{austin2021programsynthesislargelanguage}, PurpleLlama \cite{bhatt2023purplellamacybersecevalsecure}, and CyberSecEval \cite{bhatt2024cyberseceval} emphasize functional correctness or code-generation safety rather than detection accuracy. Unlike these, our work evaluates long-context retrieval and multi-label classification within a security-focused setting, offering a novel perspective on LLM behavior under complex audit-style tasks.

\section{Methodology}

To systematically evaluate LLM performance on multi-vulnerability detection, we developed a novel, automated pipeline to generate a large-scale benchmark dataset. Our methodology consists of three stages: Data Collection, Vulnerability Mapping, and Adversarial Injection.

\textbf{Data Collection.}
The foundation of our dataset consists of clean, real-world code. We utilize the github-code-dataset\cite{dataset}, filtering for C, C++, Python and JavaScript files. For each of the four target languages, we collect 1,000 unique source files.We deliberately restrict file length to 7,500–10,000 tokens. Prior work shows that LLMs exhibit significant retrieval degradation well before reaching their maximum context capacity, particularly due to positional bias and the “Lost in the Middle” phenomenon \cite{liu2023lostmiddlelanguagemodels}. This range reflects realistic file-level code encountered in real-world audits while ensuring that all evaluated models can process the full input without truncation. Importantly, this context length is sufficient to stress long-range attention and exacerbate known count and selection biases, making it a principled regime for evaluating exhaustive multi-vulnerability detection rather than mere single-bug recognition.

\textbf{Vulnerability Mapping.}
Injecting random vulnerabilities into code can result in nonsensical or implausible scenarios. To create a realistic dataset, we first determine which vulnerabilities are most suitable for a given clean file. To ensure injected vulnerabilities are syntactically and semantically feasible, we employ a "Feasibility Mapping" step. We use a high-performance LLM (Qwen2.5-32B-Instruct) as an Oracle. The model is provided with the clean source code and the \href{https://cwe.mitre.org/top25/archive/2024/2024_top25_list} {MITRE 2024 Top 25 CWE list}. It is prompted to return a JSON list of feasible CWEs that could be introduced into the code without breaking core functionality. This mapping is saved to guide the injection process.

\textbf{Adversarial Injection Pipeline}
With the mapping established, the next stage is to create the vulnerable versions of the code. We developed a VulnerabilityInjector class that operates in a hybrid mode to handle context limits.
\textbf{Input:} Clean code and a target list of CWEs (e.g., ["CWE-78", "CWE-476", ...]) based on the desired density (1, 3, 5, or 9) obtained from the mapping.
\textbf{Injection:} The model (Qwen2.5-32B-Instruct) is prompted to rewrite the code introducing all requested vulnerabilities while preserving functionality and syntactic validity of the code. The changes are asked to be made subtle and realistic along with avoiding any comments or clues that reveal the vulnerabilities.

We generated 1,000 files for each configuration (1, 3, 5, 9 vulnerabilities) across four languages, resulting in a total dataset of roughly 16,000 files.

\section{Experimental Setup}

The experimental task is defined as \textbf{multi-label vulnerability detection.} Given a code file with vulnerabilities injected, the model under evaluation must identify all CWEs present in the code from the official CWE Top 25 list. The model is expected to return a list of unique CWE IDs (e.g., ["CWE-79", "CWE-89"]).

\textbf{Models Evaluated.}
We evaluated a range of recent and powerful LLMs to assess their performance on our benchmark. The models included are: 
\textbf{Qwen2.5-32B-Instruct and 72B-Instruct}: Strong coding performance. \textbf{Llama-3.3-70B-Instruct}: State-of-the-art open-weights model. \textbf{Mistral-Small-3.2-24B}: Efficient mid-sized model. \textbf{GPT-4o-mini}: Proprietary baseline for cost-effective inference.

\textbf{Evaluation Protocol.}
We perform zero-shot prompting. The model is presented with the code and a list of definitions for the Top 25 CWEs. The system prompt explicitly instructs the model to: \textit{"Identify ALL vulnerabilities... Return ONLY the CWE IDs as a JSON array."}

\textbf{Metrics.}
To provide a holistic view of model performance, we employ standard and novel metrics to capture multi-label performance. Let P be the set of predicted CWEs and T be the set of true (ground truth) CWEs for a given file.
\begin{itemize}
    \item \textbf{Precision}: Measures the accuracy of the model's predictions: $$|P \cap T| / |P|$$. High precision indicates few false positives.
    \item \textbf{Recall}: Measures how many of the true vulnerabilities the model found: $$|P \cap T| / |T|$$. High recall indicates few false negatives.
    \item \textbf{F1-Score}: The harmonic mean of precision and recall, providing a single score for overall accuracy.
    \item \textbf{Mean Absolute Error (MAE)}: Measures the average difference between the predicted number of vulnerabilities and the actual number: $$avg(| |P| - |T| |)$$. This metric directly quantifies the model's count bias.
    \item \textbf{ExactFile Score}: A strict metric representing the fraction of files for which the model's predictions perfectly matched the ground truth (P == T).
\end{itemize}
These metrics are calculated for each file and then aggregated across each dataset to produce final scores.

\begin{table*}[t]
\centering
\small
\caption{Benchmark Results for C Across All Models and Vulnerability Levels}
\begin{tabular}{lcccccc}
\hline
\textbf{Model} & \textbf{Vulns/File} & \textbf{Precision} & \textbf{Recall} & \textbf{F1} & \textbf{MAE} & \textbf{ExactFile (\%)} \\
\hline
qwen2.5-32B & 1 & 1.0000 & 0.7024 & 0.8252 & 0.2976 & 70.2 \\
            & 3 & 1.0000 & 0.2358 & 0.3817 & 0.7642 & 7.2 \\
            & 5 & 1.0000 & 0.2134 & 0.3517 & 0.7866 & 5.5 \\
            & 9 & 1.0000 & 0.1457 & 0.2544 & 0.8544 & 2.1 \\
\hline
Qwen2.5–72B & 1 & 1.0000 & 0.9099 & 0.9528 & 0.0901 & 91.0 \\
            & 3 & 1.0000 & 0.5482 & 0.7082 & 0.4531 & 22.6 \\
            & 5 & 1.0000 & 0.4407 & 0.6118 & 0.5593 & 8.4 \\
            & 9 & 0.9997 & 0.3860 & 0.5569 & 0.6141 & 5.2 \\
\hline
Llama-3.3-70B & 1 & 0.9989 & 0.9443 & 0.9709 & 0.0557 & 94.4 \\
              & 3 & 1.0000 & 0.6215 & 0.7665 & 0.3785 & 33.6 \\
              & 5 & 1.0000 & 0.5433 & 0.7041 & 0.4567 & 16.2 \\
              & 9 & 0.9998 & 0.4638 & 0.6336 & 0.5362 & 5.1 \\
\hline
Mistral-3.2-24B & 1 & 0.9936 & 0.9484 & 0.9705 & 0.0516 & 94.8 \\
                & 3 & 1.0000 & 0.6026 & 0.7520 & 0.3974 & 25.4 \\
                & 5 & 0.9996 & 0.4982 & 0.6650 & 0.5020 & 12.9 \\
                & 9 & 1.0000 & 0.4493 & 0.6200 & 0.5507 & 4.7 \\
\hline
gpt-4o-mini & 1 & 1.0000 & 0.8988 & 0.9467 & 0.1012 & 89.9 \\
            & 3 & 1.0000 & 0.5273 & 0.6905 & 0.4727 & 28.1 \\
            & 5 & 1.0000 & 0.3974 & 0.5687 & 0.6026 & 9.2 \\
            & 9 & 0.9996 & 0.2957 & 0.4563 & 0.7044 & 2.8 \\
\hline
\end{tabular}
\end{table*}

\begin{table*}[t]
\centering
\small
\caption{Benchmark Results for C++ Across All Models and Vulnerability Levels}
\begin{tabular}{lcccccc}
\hline
\textbf{Model} & \textbf{Vulns/File} & \textbf{Precision} & \textbf{Recall} & \textbf{F1} & \textbf{MAE} & \textbf{ExactFile (\%)} \\
\hline
qwen2.5-32B & 1 & 1.0000 & 0.6420 & 0.7820 & 0.3580 & 64.2 \\
            & 3 & 0.9988 & 0.2707 & 0.4259 & 0.7294 & 12.5 \\
            & 5 & 1.0000 & 0.1870 & 0.3151 & 0.8132 & 6.4 \\
            & 9 & 1.0000 & 0.1757 & 0.2988 & 0.8243 & 4.4 \\
\hline
Qwen2.5–72B & 1 & 0.9975 & 0.8140 & 0.8965 & 0.1865 & 81.3 \\
            & 3 & 1.0000 & 0.4007 & 0.5721 & 0.5993 & 17.7 \\
            & 5 & 1.0000 & 0.2954 & 0.4561 & 0.7046 & 6.9 \\
            & 9 & 1.0000 & 0.3246 & 0.4901 & 0.6754 & 5.6 \\
\hline
Llama-3.3-70B & 1 & 1.0000 & 0.8220 & 0.9023 & 0.1780 & 82.2 \\
              & 3 & 1.0000 & 0.4383 & 0.6095 & 0.5617 & 20.9 \\
              & 5 & 1.0000 & 0.3694 & 0.5395 & 0.6306 & 9.6 \\
              & 9 & 1.0000 & 0.4494 & 0.6202 & 0.5506 & 6.7 \\
\hline
Mistral-3.2-24B & 1 & 0.7769 & 0.8950 & 0.8318 & 0.1050 & 89.5 \\
                & 3 & 1.0000 & 0.5287 & 0.6917 & 0.4713 & 27.4 \\
                & 5 & 1.0000 & 0.4546 & 0.6251 & 0.5454 & 21.5 \\
                & 9 & 0.9992 & 0.4094 & 0.5809 & 0.5907 & 8.8 \\
\hline
gpt-4o-mini & 1 & 1.0000 & 0.8020 & 0.8901 & 0.1980 & 80.2 \\
            & 3 & 1.0000 & 0.3410 & 0.5086 & 0.6590 & 17.3 \\
            & 5 & 1.0000 & 0.2458 & 0.3946 & 0.7542 & 9.6 \\
            & 9 & 1.0000 & 0.2392 & 0.3861 & 0.7608 & 5.4 \\
\hline
\end{tabular}
\end{table*}

\begin{table*}[t]
\centering
\small
\caption{Benchmark Results for JavaScript Across All Models and Vulnerability Levels}
\begin{tabular}{lcccccc}
\hline
\textbf{Model} & \textbf{Vulns/File} & \textbf{Precision} & \textbf{Recall} & \textbf{F1} & \textbf{MAE} & \textbf{ExactFile (\%)} \\
\hline
qwen2.5-32B & 1 & 1.0000 & 0.7770 & 0.8745 & 0.2230 & 77.7 \\
            & 3 & 1.0000 & 0.4100 & 0.5816 & 0.5900 & 18.5 \\
            & 5 & 1.0000 & 0.3213 & 0.4864 & 0.6787 & 10.2 \\
            & 9 & 1.0000 & 0.2261 & 0.3688 & 0.7739 & 3.8 \\
\hline
Qwen2.5–72B & 1 & 0.9946 & 0.9230 & 0.9575 & 0.0775 & 92.2 \\
            & 3 & 1.0000 & 0.5213 & 0.6854 & 0.4787 & 22.6 \\
            & 5 & 1.0000 & 0.4080 & 0.5796 & 0.5924 & 10.7 \\
            & 9 & 0.9996 & 0.3088 & 0.4718 & 0.6913 & 5.2 \\
\hline
Llama-3.3-70B & 1 & 0.9873 & 0.9300 & 0.9578 & 0.0700 & 93.0 \\
              & 3 & 1.0000 & 0.6077 & 0.7560 & 0.3923 & 30.6 \\
              & 5 & 1.0000 & 0.5031 & 0.6694 & 0.4969 & 13.9 \\
              & 9 & 1.0000 & 0.4323 & 0.6037 & 0.5678 & 6.6 \\
\hline
Mistral-3.2-24B & 1 & 0.9897 & 0.9580 & 0.9736 & 0.0420 & 95.8 \\
                & 3 & 1.0000 & 0.6030 & 0.7523 & 0.3970 & 26.8 \\
                & 5 & 1.0000 & 0.4867 & 0.6547 & 0.5133 & 14.4 \\
                & 9 & 0.9997 & 0.4052 & 0.5767 & 0.5948 & 5.2 \\
\hline
gpt-4o-mini & 1 & 1.0000 & 0.9480 & 0.9733 & 0.0520 & 94.8 \\
            & 3 & 1.0000 & 0.5500 & 0.7097 & 0.4500 & 26.5 \\
            & 5 & 1.0000 & 0.3920 & 0.5632 & 0.6080 & 11.3 \\
            & 9 & 1.0000 & 0.2669 & 0.4213 & 0.7332 & 5.1 \\
\hline
\end{tabular}
\end{table*}

\begin{table*}[t]
\centering
\small
\caption{Benchmark Results for Python Across All Models and Vulnerability Levels}
\begin{tabular}{lcccccc}
\hline
\textbf{Model} & \textbf{Vulns/File} & \textbf{Precision} & \textbf{Recall} & \textbf{F1} & \textbf{MAE} & \textbf{ExactFile (\%)} \\
\hline
qwen2.5-32B & 1 & 1.0000 & 0.4740 & 0.6431 & 0.5260 & 47.4 \\
            & 3 & 1.0000 & 0.3127 & 0.4764 & 0.6873 & 17.7 \\
            & 5 & 1.0000 & 0.3972 & 0.5686 & 0.6044 & 20.8 \\
            & 9 & 1.0000 & 0.4464 & 0.6173 & 0.5538 & 12.5 \\
\hline
Qwen2.5–72B & 1 & 0.9701 & 0.4860 & 0.6476 & 0.5140 & 48.6 \\
            & 3 & 0.9981 & 0.3577 & 0.5266 & 0.6425 & 21.1 \\
            & 5 & 1.0000 & 0.4200 & 0.5915 & 0.5820 & 13.8 \\
            & 9 & 0.9998 & 0.4940 & 0.6613 & 0.5064 & 13.1 \\
\hline
Llama-3.3-70B & 1 & 0.9093 & 0.5110 & 0.6543 & 0.4941 & 50.5 \\
              & 3 & 0.9992 & 0.4013 & 0.5727 & 0.5988 & 22.9 \\
              & 5 & 1.0000 & 0.5062 & 0.6722 & 0.4938 & 17.9 \\
              & 9 & 0.9998 & 0.5938 & 0.7451 & 0.4063 & 18.9 \\
\hline
Mistral-3.2-24B & 1 & 0.9936 & 0.9484 & 0.9705 & 0.0516 & 94.8 \\
                & 3 & 1.0000 & 0.6026 & 0.7520 & 0.3974 & 25.4 \\
                & 5 & 0.9996 & 0.4982 & 0.6650 & 0.5020 & 12.9 \\
                & 9 & 1.0000 & 0.4493 & 0.6200 & 0.5507 & 4.7 \\
\hline
gpt-4o-mini & 1 & 1.0000 & 0.5530 & 0.7122 & 0.4470 & 55.3 \\
            & 3 & 1.0000 & 0.3767 & 0.5472 & 0.6233 & 24.2 \\
            & 5 & 1.0000 & 0.4790 & 0.6477 & 0.5210 & 24.3 \\
            & 9 & 1.0000 & 0.5038 & 0.6701 & 0.4962 & 18.3 \\
\hline
\end{tabular}
\end{table*}

\section{Results}
Our experiments reveal that while modern LLMs have improved at detecting isolated faults, their reliability degrades significantly when confronted with complex, multi-vulnerability files.

\subsection{Performance Overview Across Languages}
The transition from single to multi-vulnerability detection caused a universal performance degradation across all models.
We observe that C and JavaScript are generally "easier" for models to audit than C++ and Python in single-vulnerability settings.
\textbf{C} (Llama-3.3-70B): 94.4\% Recall (1-vuln), \textbf{JavaScript} (Llama-3.3-70B): 93.0\% Recall (1-vuln),\textbf{C++} (Llama-3.3-70B): 82.2\% Recall (1-vuln) and \textbf{Python} (Llama-3.3-70B): 51.1\% Recall (1-vuln).

This discrepancy suggests that models may struggle with the specific syntactic structures or vulnerability patterns common in Python and C++ compared to the more explicit flaw patterns often found in C and JavaScript.

\subsection{Compiled vs Interpreted Languages}

We observed a notable divergence in performance trends between compiled languages (C, C++) and interpreted languages (Python, JavaScript), suggesting that model architecture or training data distribution influences how these distinct semantic contexts are processed.

\textbf{Compiled Languages (C \& C++):}
In C and C++, models displayed a linear degradation pattern. Performance was strongest on files with a single vulnerability but dropped precipitously as density increased.
\begin{itemize}
    \item \textbf{C}: This was the "easiest" language for the top model, Llama-3.3-70B, which achieved a 94.4\% Recall on single-vulnerability files. However, this dropped to 46.4\% on 9-vulnerability files—a nearly 50\% reduction in effectiveness.
    \item \textbf{C++}: This language proved harder overall. Qwen2.5-72B started with 81.4\% Recall (1-vuln) but collapsed to 24.6\% Recall for 5-vuln files. The complex syntax and memory management features of C++ likely contribute to this difficulty, increasing the "cognitive load" on the model's context window.
\end{itemize}

\textbf{Interpreted Languages (Python \& JavaScript):}
Interpreted languages showed higher variance and, in the case of Python, an anomalous "inverted" trend.
\begin{itemize}
    \item \textbf{JavaScript}: Similar to C, JavaScript performance degraded linearly. Mistral-3.2-24B showed excellent initial robustness (95.8\% Recall for 1-vuln) but fell to 40.9\% for 9-vulns.
    \item \textbf{Python}: Python performance was unique. Llama-3.3-70B started with a relatively low 51.1\% Recall for single flaws but improved to 59.4\% Recall on 9-vulnerability files. This suggests that Python vulnerabilities (often related to dynamic typing or library misuse) might trigger stronger association networks in the model when multiple flaws are present, or that the model's "vulnerability threshold" for Python is calibrated differently, leading to more aggressive predictions in dense files.
\end{itemize}

\begin{figure*}[t]
  \includegraphics[width=0.96\linewidth]{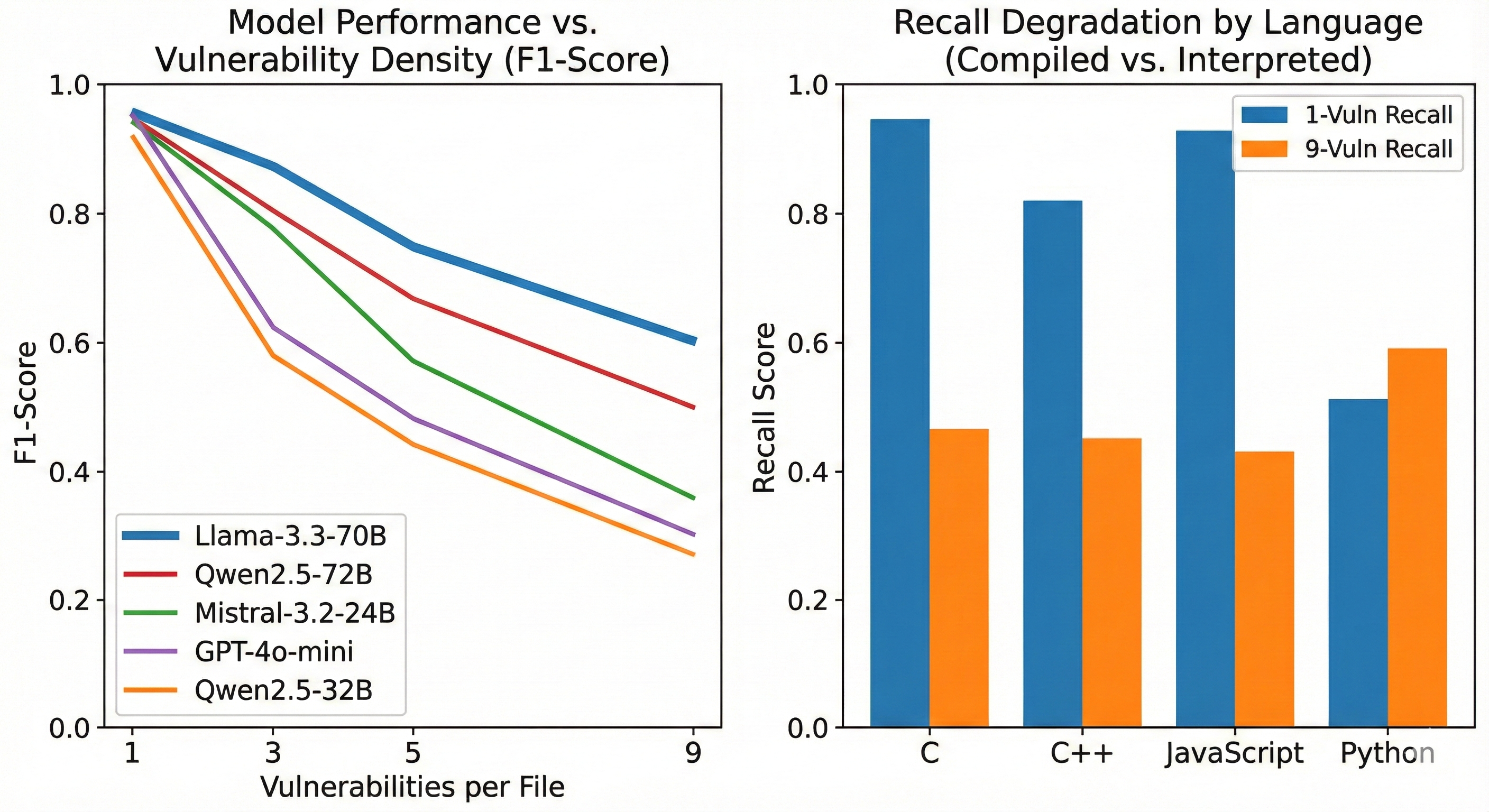} 
  \caption {Evaluation of LLMs on Code Vulnerability Detection Across Varying Density and Language Types.}
\end{figure*}

\subsection{Count Bias \& Recall Analysis}

A primary finding of this study is the prevalence of Count Bias—the tendency of LLMs to under-report the total number of vulnerabilities, effectively "giving up" after finding a few issues.

\textbf{Recall Degradation}: Across all languages (except Python), the recall curve flattens as vulnerability count ($N$) increases. Ideally, recall should remain constant (finding 90\% of 1 issue $\approx$ finding 90\% of 9 issues). Instead, we see that for N=9, models typically retrieve only 3–4 vulnerabilities (Recall $\approx$ 0.3–0.4).

\textbf{ExactFile Failure}: The ExactFile metric (perfectly identifying all $N$ vulnerabilities with zero false positives) highlights the severity of this bias. For Qwen2.5-32B on C++, the ExactFile score dropped from 64.2\% (1-vuln) to just 4.4\% (9-vulns). This indicates that in a real audit of a highly flawed file, this model would provide an incomplete report 95% of the time.

\textbf{MAE} (Mean Absolute Error): The MAE for vulnerability counts consistently increased with $N$. On 9-vuln C files, Qwen2.5-32B had a normalized MAE of 0.85, meaning its estimated count was often off by nearly one full standard deviation of the possible range.

\subsection{Model Comparison}

We benchmarked models of varying sizes to determine if parameter count correlates with multi-vulnerability auditing capability.

\textbf{Llama-3.3-70B (Top Performer)}: This model was the most robust overall. It consistently maintained the highest recall in dense files (e.g., 46\% Recall on 9-vuln C files vs. 24\% for Qwen2.5-72B). It appears less susceptible to count bias, likely due to better instruction-following capabilities regarding exhaustive search.

\textbf{Qwen2.5-72B}: While strong on single-vulnerability tasks (often matching Llama-3.3), it suffered from steeper degradation. On C++, its recall dropped by ~57\% when moving from 1 to 5 vulnerabilities, compared to a shallower drop for Llama-3.3.

\textbf{Mistral-3.2-24B}: Surprisingly efficient for its size. On JavaScript, it achieved the highest single-vulnerability recall (95.8\%) of any model tested, outperforming even the 70B parameter models. However, its small context capacity likely contributed to its rapid decline in the 9-vulnerability setting (40.9\% Recall).

\textbf{GPT-4o-mini}: This model performed competitively on simpler tasks (e.g., 94.8\% Recall on 1-vuln JavaScript) but struggled with density. It had the lowest recall on the 9-vuln JavaScript dataset (26.7\%), indicating that smaller, optimized commercial models may prioritize precision and speed over the exhaustive context processing needed for deep audits.

\section{Conclusion \& Future Work}

In this work, we introduced the first large-scale benchmark designed explicitly to evaluate Large Language Models on \textbf{multi-vulnerability detection} at the file level, moving beyond traditional single-bug or binary classification tasks. Our dataset, spanning four languages and controlled vulnerability densities, revealed that even state-of-the-art models struggle significantly when required to identify all vulnerabilities within long-context code. Despite strong performance on single-vulnerability scenarios, every evaluated model, even Llama-3.3-70B, exhibited substantial degradation in recall, rising count bias, and sharp declines in ExactFile accuracy as the number of vulnerabilities increased. These findings highlight a fundamental limitation of current LLM architectures: while they excel at surface-level pattern recognition, they remain unreliable for comprehensive security audits involving complex, dense, real-world codebases.

Looking ahead, several promising research directions emerge. First, future work should explore \textbf{hybrid static analysis + LLM systems} to mitigate count bias and ensure exhaustive search through structural cues rather than purely generative reasoning. Second, extending the benchmark to include naturally occurring vulnerabilities, broader CWE categories, and logic-level flaws would improve ecological validity. Additionally, architectural advances, such as retrieval-augmented long-context reasoning, modular verification steps, or specialized multi-pass “exhaustive audit” prompting, may help LLMs better handle dense vulnerability clusters. Finally, incorporating human-in-the-loop evaluation and studying model behavior under adversarial prompts can deepen understanding of failure modes. We hope this benchmark serves as a foundation for more realistic and rigorous evaluation of LLM-based security systems and catalyzes progress toward trustworthy automated vulnerability detection.

\section{Limitations}

\textbf{Synthetic Data Generation}: While we employed an Oracle LLM to ensure feasibility, the vulnerabilities are injected into existing code rather than occurring "in the wild." This may create artifacts (e.g., sudden style changes) that models might latch onto.

\noindent \textbf{Scope of Vulnerabilities}: We limited the scope to the MITRE 2024 Top 25 CWEs. Real-world code contains domain-specific logic flaws that fall outside these categories.

\noindent \textbf{Ground Truth Reliance}: Our evaluation relies exclusively on the mapping of induced vulnerabilities as the ground truth. We assume the pre-injection source code, sourced from GitHub, is free of vulnerabilities. This creates a limitation where valid detections of pre-existing bugs in the original code are incorrectly penalized as False Positives, potentially masking the true detection capabilities of the models on legacy codebases.

% \section{References}

% \begin{itemize}
%     \item Austin, J., et al. (2021). "Program Synthesis with Large Language Models." arXiv preprint arXiv:2108.07732.
%     \item Chakraborty, S., et al. (2021). "Deep Learning based Vulnerability Detection: Are We There Yet?" IEEE Transactions on Software Engineering.
%     \item Chen, M., et al. (2021). "Evaluating Large Language Models Trained on Code." arXiv preprint arXiv:2107.03374.
%     \item Liu, N. F., et al. (2023). "Lost in the Middle: How Language Models Use Long Contexts." arXiv preprint arXiv:2307.03172.
%     \item Meta. (2023). "Purple Llama: CyberSecEval."
%     \item Pezeshkpour, P., et al. (2025). "Large Language Models Sensitivity to The Order of Options in Multiple-Choice Questions." arXiv preprint arXiv:2506.00643.
%     \item Zhou, Y., et al. (2019). "Devign: Effective Vulnerability Identification by Learning Comprehensive Program Semantics via Graph Neural Networks." NeurIPS.
%     \item Xu, W., et al, (2025). "SATA-BENCH: Select all that Apply for Multiple Choice Question." arXiv:2506.00643
% \end{itemize}

\bibliography{latex/custom}

\end{document}